\documentclass[english,11pt,oneside]{article}
\usepackage{amsmath}
\usepackage{authblk}
\usepackage{amsfonts}
\usepackage{caption}
\usepackage{amssymb}
\usepackage{todonotes}
\usepackage{graphicx, rotating}
\usepackage{epstopdf}
\usepackage{afterpage}
\usepackage{epsfig}
\usepackage{latexsym}
\usepackage{multirow}
\usepackage{color}
\usepackage{slashed}
\usepackage{pifont}
\usepackage{comment}
\usepackage{array}
\usepackage{orcidlink}
\usepackage{cite}
\usepackage{soul}
\usepackage[top=2.8 cm, bottom=2.8 cm, left=3 cm, right=3 cm]{geometry}

\usepackage{colortbl}
\usepackage{booktabs}
\usepackage{xspace}
\usepackage{xcolor}
\usepackage{chngpage}
\usepackage{upquote}
\usepackage{mathtools}

\usepackage[title]{appendix}

\newcommand{\dd}{\text{d}}

\newcommand{\tr}{\mathrm{tr}}
\newcommand{\ba}{\begin{align}}
\newcommand{\ea}{\end{align}}
\newcommand{\no}{\nonumber}

\newcommand{\be}{\begin{equation}}
\newcommand{\ee}{\end{equation}}
\newcommand{\bea}{\begin{eqnarray}}
\newcommand{\eea}{\end{eqnarray}}
\newcommand{\bit}{\begin{itemize}}
\newcommand{\eit}{\end{itemize}}

\usepackage{enumitem}

\usepackage{hyperref}
\hypersetup{
 colorlinks = true,
 linkcolor = red,
 citecolor={blue},
urlcolor = blue
}

\usepackage{bbold}

\begin{document}

\title{{\huge{
Spurious gauge-invariance and $\gamma_5$\\ in Dimensional Regularization}}
%Spurious gauge-invariance in the BMHV scheme
%Restoring gauge-invariance in Dimensional Regularization
%Dimensional regularization and (spurious) gauge-invariance
}

\author[a\,b]{Pablo Olgoso Ruiz %\orcidlink{...},
\thanks{pablo.olgosoruiz@unipd.it}}
\author[b]{Luca Vecchi 
%\orcidlink{0000-0001-5254-8826},
\thanks{luca.vecchi@pd.infn.it}}

\affil[a]{\small\emph{Dipartimento di Fisica e Astronomia, Universit\`a di Padova, Via F. Marzolo 8, 35131 Padova, Italy}}

\affil[b]{\small\emph{Istituto Nazionale di Fisica Nucleare, Sezione di Padova, Via F. Marzolo 8, 35131 Padova, Italy}}

\date{}

\maketitle
\thispagestyle{empty}

\begin{abstract}

Dimensional regularization is arguably the most popular and efficient scheme for multi-loop calculations. Yet, when applied to chiral (gauge) theories like the Standard Model and its extensions, one is forced to deal with the infamous ``$\gamma_5$ problem". The only formulation that has been demonstrated to be consistent at all orders in perturbation theory, known as Breiteinlhoner-Maison-'t Hooft-Veltman scheme, is rather cumbersome because of the lack of manifest chiral gauge-invariance. In this paper we point out that this drawback can be alleviated by the introduction of auxiliary fields that restore a spurious version of gauge-invariance. If combined with the background field method, all 1PI amplitudes and the associated counterterms are formally covariant and thus severely constrained by the symmetries. As an illustration we evaluate the symmetry-restoring counterterms at 1-loop in the most general renormalizable gauge theory with Dirac fermions and scalar fields, the Standard Model representing a particular example.

\end{abstract}

\newpage

{\hypersetup{linkcolor=black}	\tableofcontents}	

%\newpage

\section{Introduction}
\label{sec:introduction}

Dimensional regularization (Dim-Reg) \cite{tHooft:1972tcz, Bollini:1972ui} is a very popular scheme in perturbative calculations. This is so for very good reasons: Dim-Reg is compatible with vector-like gauge symmetries and Lorentz invariance, it simultaneously regulates both UV and IR divergences, and it is straightforwardly applicable to higher orders. For all these reasons Dim-Reg has by now become the standard regularization scheme for multi-loop calculations.

Already at its conception, however, 't Hooft and Veltman noted a complication in dealing with chiral theories like the Standard Model \cite{tHooft:1972tcz}. At the heart of Dim-Reg is the fact that convergence of the loop integrals is ensured by extending the kinetic terms of all fields to $d$-dimensions. Yet, in $d$-dimensions there is no notion of chirality and hence chiral symmetries are inevitably broken.

In the approach of 't Hooft and Veltman, and subsequently formalized by Breitenlohner and Maison \cite{Breitenlohner:1977hr}, the Dirac gamma matrices are assumed to satisfy the $d$-dimensional Clifford algebra and split into a four-dimensional set ($\gamma^{\bar\mu}$) and a $d-4$-dimensional set ($\gamma^{\hat\mu}$). Instead, the $\gamma_5$ matrix and the Levi-Civita tensor remain genuinely four-dimensional objects and satisfy (we use $\epsilon_{0123}=1$)
\be\label{gamma5def}
\gamma_5=\frac{i}{4!}\epsilon_{\bar\mu\bar\nu\bar\alpha\bar\beta}\gamma^{\bar\mu}\gamma^{\bar\nu}\gamma^{\bar\alpha}\gamma^{\bar\beta}.
\ee
From these definitions follows that $\gamma_5$ anti-commutes only with the four-dimensional gamma matrices, but commutes with the others
\be\label{anticommcomm}
\left\{\gamma^{\bar\mu},\gamma_5\right\}=0,~~~~~~~\bigl[\gamma^{\hat\mu},\gamma_5\bigr]=0.
\ee
As a result of \eqref{anticommcomm}, the $d$-dimensional kinetic term of fermions mixes left and right chiralities, violating all chiral symmetries, global and local (see \cite{Belusca-Maito:2023wah} for a review and more references). Eq. \eqref{gamma5def}, and its consequence \eqref{anticommcomm}, define what we will refer to as the {\emph{BMHV scheme}}.

The breaking of the chiral symmetries implied by the BMHV scheme is of course only fictitious and has no physical consequences, if properly handled. Consider for instance a theory with a local chiral symmetry. A regularization-independent condition guaranteeing that our theory has no gauge anomalies can be formulated as a  requirement on the fermionic generators. Specifically, denoting by $T_{L,R}^A$ the gauge generators of the left and right chiralities, a well-defined (anomaly-free) gauge theory must satisfy  
\be\label{D=0}
D^{ABC}=\tr\left(T^A_L\left\{T^B_L,T^C_L\right\}\right)-\tr\left(T^A_R\left\{T^B_R,T^C_R\right\}\right)=0.
\ee
When \eqref{D=0} holds, if the regularization scheme violates the gauge symmetry one can always add appropriate {\emph{gauge non-invariant}} counterterms that eliminate unphysical degrees of freedom order by order in perturbation theory. Those counterterms must be included in addition to the more familiar ones that remove the $1/(d-4)$ poles, which in such non-invariant schemes also contain gauge non-invariant pieces in general. In an anomaly-free gauge theory satisfying \eqref{D=0}, therefore, the BMHV scheme is a perfectly viable one; it just requires more work. In the last years the explicit form of the BRST-restoring counterterms for various chiral gauge theories regularized via the BMHV scheme has been calculated at the first non-trivial orders in  \cite{Martin:1999cc,Sanchez-Ruiz:2002pcf,Belusca-Maito:2020ala,Belusca-Maito:2021lnk,Belusca-Maito:2022wem,Stockinger:2023ndm}. Similarly, the counterterms necessary to restore gauge-invariance when using the background field method have been calculated for various theories at the one-loop order in \cite{Cornella:2022hkc,Naterop:2023dek}.

Even though the fictitious violation of chiral (global or local) invariance induced by the BMHV scheme is not a conceptual obstacle, it is certainly a practical nuisance. Bare correlators do not satisfy the generalized Ward identities prior to the introduction of the symmetry-restoring counterterms, and so a priori there is no manifest guiding principle that controls the structure of the radiative corrections. This lack of symmetry complicates calculations especially in chiral gauge theories like the electroweak sector of the Standard Model. For this reason, alternative approaches to the ``$\gamma_5$ problem" in Dim-Reg have been proposed in the last decades with the goal of limiting as much as possible the regularization-induced violation of chiral symmetries. Particularly popular are the so-called ``naive Dim-Reg" \cite{Chanowitz:1979zu} and ``KKS" \cite{Kreimer:1989ke,Korner:1991sx} schemes (see also \cite{Jegerlehner:2000dz}). Yet, none of those alternative approaches rely on rigorous mathematical definitions of $\gamma_5$, and as such are inherently ambiguous.~\footnote{A concrete indication of possible shortcomings of the KKS scheme is found in the evaluation of the ABJ anomaly at 2-loops \cite{Chen:2023lus} or in the beta functions of the Standard Model couplings at 4-loops \cite{Bednyakov:2015ooa, Zoller:2015tha}.} Nevertheless, the vast majority of groups in our community still favors the technical simplicity of the alternatives over the cumbersome rigour of the BMHV scheme.

In this paper we will adopt the BMHV scheme for $\gamma_5$, the only approach demonstrated to be fully consistent at all orders in perturbation theory. Our goal is to find some ``order" in the apparent chaos associated to that formalism. Specifically, we will see that the structure of the counterterms and the correlators themselves are constrained by the way Dim-Reg explicitly breaks the chiral symmetry: a {\emph{spurious gauge-invariance}} will guide our calculation and replace the broken gauge (or BRST) invariance.

We begin in Section \ref{sec:standard} with an introduction of the standard regularization procedure. This preliminary discussion is a summary of known results and mostly serves the purpose of introducing our notation. A version of the regularized theory that preserves the spurious chiral invariance is introduced in Section \ref{sec:spurion}. In Section \ref{sec:radcorr} we discuss a few of the advantages deriving from the spurious invariance. As a concrete application we identify the complete set of symmetry-restoring counterterms for an arbitrary renormalizable gauge theory with Dirac fermions and scalars at 1-loop, generalizing the earlier work of \cite{Cornella:2022hkc}. The case of the Standard Model can be obtained using the expressions in Appendix \ref{app:SM}. In Appendix \ref{sec:Lorentz} we point out that, if desired, one can build a regularized theory that also respects a spurious version of $d$-dimensional Lorentz invariance. We conclude in Section \ref{sec:conclusion}.

\section{Regularized theory}
\label{sec:reg}

In this section we introduce the regularized action for a chiral gauge theory with charged fermionic fields $\Psi_i$, where $i,j,\dots$ collectively denote both gauge and flavor indices, and scalars $\phi_a$, with $a,b,\dots$ again running over both gauge and flavor indices. The gauge symmetry is assumed to be the direct product of simple and abelian unitary groups, but is otherwise completely arbitrary. The representations of the fermions and scalars are in general reducible. We describe fermions via a Dirac field $\Psi$ because more suitable for Dim-Reg; theories of Weyl fermions may be obtained via the introduction of unphysical dummy components of $\Psi$, that are effectively decoupled in four dimensions. The scalar fields $\phi_a$ are taken to be real without loss of generality.

\subsection{The standard theory}
\label{sec:standard}

As reviewed in Section \ref{sec:introduction}, in Dim-Reg the kinetic terms of all fields are extended to $d$ dimensions. In the case of bosons this can be trivially achieved while preserving all symmetries, whereas some subtlety arises when considering chiral fermions, as we will review shortly. 

The regularized action is the sum of various components:
\be\label{LagTot}
S^{\text{Reg}}_{\text{std}}[\xi]
\equiv\int \dd^dx\left[{\cal L}^{\text{Reg}}_{\text{Bos}}+{\cal L}^{\text{Reg}}_{\text{Yuk}}+{\cal L}_{\text{Kin}}^{\text{Reg}}\right],
\ee
with $\xi$ collectively denoting all fields. The first one involves the bosonic degrees of freedom and includes the kinetic terms of the gauge fields and the scalars as well as the scalar potential:~\footnote{We use mostly-minus signature.} 
\be\label{bosons}
{\cal L}^{\text{Reg}}_{\text{Bos}}=-\frac{1}{4}F^A_{\mu\nu}F^{A\,\mu\nu}+(D_\mu\phi)_a^\dagger (D^\mu\phi)_a-V(\phi).
\ee
These terms are obtained by simply promoting the four-dimensional theory to a $d$-dimensional one, so $\mu$ runs over both the four-dimensional coordinates $\bar\mu=0,1,2,3$ and the evanescent ones $\hat\mu$. We define $D_\mu\phi_a=\partial_\mu\phi_a-iA^A_\mu [T_\phi^A]_{ab}\phi_b$, with $T_\phi^A$ the (hermitian and purely imaginary) generators of the local transformations of $\phi$.

The second and third terms in \eqref{LagTot} involve the fermionic fields $\Psi$. Before writing them explicitly let us make some preliminary considerations. In four dimensions the left and right fermion chiralities form independent Lorentz components and can thus be rotated independently. In that case the fermionic kinetic term possesses a large $SU(N_f)\times SU(N_f)$ global symmetry (in general broken by the weak gauging and the Yukawa interactions) and the left and right chiralites may carry different representations under the gauge group, generated by hermitian matrices $T_L^A$ and $T_R^A$ respectively. Ideally, we would like to retain the same features also in $d$-dimensions. Thus we define the projectors 
\be\label{PLPR}
P_L=\frac12(1-\gamma_5),~~~~~~P_R=\frac12(1+\gamma_5)
\ee
analogously to what done in four dimensions, but here in terms of the object $\gamma_5$ given in \eqref{gamma5def}. Conventionally we still call $P_L\Psi$ the ``left" and $P_R\Psi$ the ``right" chiralities,  respectively. However, these are no more independent fields in $d$-dimensions. Nevertheless, in the basis of the orthogonal subspaces defined by $P_L$, $P_R$, we can postulate that under a $d$-dimensional gauge rotation fermions transform as
\be\label{gaugetrans}
\Psi_i\to U_{ij}\Psi_j,~~~~~~\overline\Psi_i\to {\overline\Psi}_j\, [\overline U^\dagger]_{ji}, 
\ee
where
\be
U=e^{i\alpha^AT^A}%=\left(\begin{matrix}e^{i\alpha^AT_L^a} & 0\\ 0 & e^{i\alpha^AT_R^a}\end{matrix}\right)
,~~~~~~\overline U=\gamma^0U\gamma^0=e^{i\alpha^A\overline T^A}%=\left(\begin{matrix}e^{i\alpha^AT_R^a} & 0\\ 0 & e^{i\alpha^AT_L^a}\end{matrix}\right).
\ee
with $\alpha^A$ the gauge parameter and%~\footnote{By construction $UT^AU^\dagger=R^{ab}T^b$ and $\overline U\,\overline T^A\overline U^\dagger=R^{ab}\overline T^b$, where $R^{ab}$ are the same real matrices.}
\be
T^A=P_LT^A_L+P_RT^A_R,~~~~~~\overline T^A=\gamma^0T^A\gamma^0=P_RT^A_L+P_LT^A_R.
\ee

Having clarified what we mean by chiral gauge transformations in $d$ dimensions, we can now have a look at the fermionic interactions in \eqref{LagTot}. We start with the second term, which includes the couplings between fermions and scalars. For definiteness we assume that the global fermion number is conserved and focus on Dirac-type interactions. Our results can be straightforwardly generalized to Majorana-type Yukawas, but we feel no urgency to do so at present given that no technical complication would emerge in that case and, moreover, the Standard Model belongs to the class of scenarios we consider. We define
\be\label{Yukawa}
{\cal L}^{\text{Reg}}_{\text{Yuk}}=-Y^a_{ij}\overline{\Psi_i}P_R\Psi_j\phi_a+{\rm h.c.},
\ee
where gauge-invariance in $d$-dimensions is ensured if
\be
[T^A_L]_{ik}Y^a_{kj}-Y^a_{ik}[T^A_R]_{kj}=Y^b_{ij}[T^A_\phi]_{ba}.
\ee
The structure of the regularized Yukawa operator is formally the same as in four dimensions. The story, however, complicates when considering the fermionic kinetic term.

A natural generalization of the Dirac Lagrangian to $d$ dimensions leads to
\begin{align}\label{Reg1}
{\cal L}_{\text{Kin}}^{\text{Reg}}=\overline{\Psi}i\gamma^{\bar\mu}D_{\bar\mu}\Psi+\frac{1}{2}\left\{\overline{\Psi}\gamma^{\hat\mu} i\left(\partial_{\hat\mu}-icA^A_{\hat\mu}T^A\right)\Psi+{\rm h.c.}\right\},
\end{align}
where $D_\mu\Psi=(\partial_\mu-iA^A_\mu T^A)\Psi$ and $c$ is an arbitrary number (more on this below). Crucially, \eqref{Reg1} inevitably breaks the chiral symmetries. Indeed, under \eqref{gaugetrans} the fermion current transforms as
\be\label{breaking}
\overline{\Psi}\gamma^{\mu}\Psi\to \overline{\Psi}\,\overline U^\dagger \gamma^{\mu}U\Psi
=\begin{cases}
    \overline{\Psi} \gamma^{\bar\mu}\Psi & {\text{for}}~\mu=\bar\mu\\
    \overline{\Psi}\,\overline U^\dagger U\gamma^{\hat\mu}\Psi& {\text{for}}~\mu=\hat\mu,
\end{cases}
\ee
where in the last step we used \eqref{anticommcomm}, according to which the gauge matrices satisfy $\gamma^{\hat\mu}U=U\gamma^{\hat\mu}$ and $\gamma^{\bar\mu}U=\overline U\gamma^{\bar\mu}$. Therefore, the evanescent part of the kinetic term in \eqref{Reg1} cannot be invariant if $\overline U\neq U$, namely if the transformation is chiral. The problem obviously applies to both global and local chiral symmetries.

Because chiral invariance is anyway lost, we can appreciate the meaning of the parameter $c$ introduced in \eqref{Reg1}. For $c=1$, the evanescent derivative in the brackets of \eqref{Reg1} is covariant under the entire $d$-dimensional gauge rotations, i.e. under local transformations \eqref{gaugetrans} with arbitrary $\alpha^A(x)$. Yet, since the chiral symmetry is broken there is no reason to prefer $c=1$ over any other choice. Hence the value of $c$ can be taken to be arbitrary, it merely defines a scheme-dependence. We will work with $c=0$ in our explicit calculations.

To conclude, the action given in \eqref{LagTot}, with the three components shown in \eqref{bosons}, \eqref{Yukawa} and \eqref{Reg1}, is the dimensionally regularized version of the theory we are interested in. For this reason we will refer to it as the {\emph{standard}} regularized theory. Regrettably, it is not invariant under chiral transformations because of the evanescent part in \eqref{Reg1}. Hence one is forced to introduce symmetry-restoring counterterms order by order to ensure the theory satisfies the generalized Ward identities for chiral symmetries, as already anticipated below Eq. \eqref{D=0}. There is instead no need of symmetry-restoring counterterms for vector-like internal symmetries, as these are preserved by our regularization.

It is also important to stress that the presence of the chirality projectors in  \eqref{Yukawa} and \eqref{Reg1} breaks $d$-dimensional Lorentz invariance. Only the four-dimensional subgroup $O(1,3)$ and rotations $O(d-4)$ in the extra dimensions are preserved. This is a consequence of the fact that $\gamma_5$ is not covariant since the $\bar\mu\hat\nu$ components of the Lorentz generators do not commute with it. But this is not a serious problem in practice because we are ultimately interested in the four-dimensional physics, and our regularized Lagrangian is invariant under the four-dimensional Lorentz symmetry. There is thus no need to add Lorentz-restoring counterterms. The only drawback is that the bare correlator functions derived from \eqref{LagTot} are not $d$-dimensional Lorentz covariant but rather $O(1,3)\times O(d-4)$ tensors.

\subsection{Spurious chiral-invariance}
\label{sec:spurion}

The violation of the chiral symmetries appearing in \eqref{breaking} is somehow reminiscent of the explicit breaking of the chiral $SU(N_f)\times SU(N_f)$ symmetry of QCD induced by a quark mass proportional to the identity. In that context a way to describe low-energy QCD while still enjoying the full power of $SU(N_f)\times SU(N_f)$ is well known: one introduces an appropriate Nambu-Goldstone matrix and ``dresses" the quark mass term with them. This way one can interpret the quark mass as a field transforming covariantly under chiral transformations and the full chiral symmetry is formally restored. This trick proves extremely useful when analyzing the dependence of physical observables on quark masses.

Borrowing from that experience we introduce in our theory an auxiliary, non-dynamical scalar field matrix $\Omega$ transforming as 
\be\label{defOmega}
\Omega\to e^{i\alpha^AT_L^A}\Omega e^{-i\alpha^AT_R^A}
\ee
under the gauge symmetry (or analogously for global chiral rotations). With $\Omega$ at our disposal, we can formally recover invariance under the full chiral symmetry and retain control over the symmetry-breaking consequences of our regularization. It is sufficient to introduce the combination 
\be\label{transfO}
(\Omega^\dagger P_L+\Omega P_R)\to \overline U(\Omega^\dagger P_L+\Omega P_R)  U^\dagger
\ee
in the evanescent component of the fermion kinetic term and replace the regularized action \eqref{LagTot} with
\begin{align}\label{Action}
S^{\text{Reg}}[\xi,\Omega]
&\equiv\int \dd^dx\left[{\cal L}_{\text{Bos}}^{\text{Reg}}+{\cal L}_{\text{Yuk}}^{\text{Reg}}\right]\\\no
&+\frac12\int\dd^dx\left\{\overline{\Psi}\gamma^{\bar\mu}iD_{\bar\mu}\Psi+\overline{\Psi}\gamma^{\hat\mu}\left(\Omega^\dagger P_L+\Omega P_R\right) i\left(\partial_{\hat\mu}-icA^A_{\hat\mu}T^A\right)\Psi+{\rm h.c.}\right\}.
\end{align}
The non-symmetric standard theory of Section \ref{sec:standard} is a particular case of \eqref{Action} in which the auxiliary field is set to a background equal to the identity matrix 
\be\label{physback}
\Omega\to{\mathbb 1},
\ee
namely $S^{\text{Reg}}[\xi,{\mathbb 1}]=S_{\text{std}}^{\text{Reg}}[\xi]$. Yet, when $\Omega$ is turned on, \eqref{Action} is formally invariant under {\emph{spurious}} $d$-dimensional gauge transformations (namely gauge transformations with $\alpha^A=\alpha^A(x^{\mu})$ functions of $x^{\mu}$) if $c=1$, or only invariant under the {\emph{spurious}} four-dimensional restriction (namely gauge transformations with $\alpha^A=\alpha^A(x^{\bar\mu})$ functions of $x^{\bar\mu}$ alone) if $c\neq1$. The local rotation of the dynamical fields is exactly compensated by the rotation \eqref{defOmega} of the auxiliary field $\Omega$. For this reason we will refer to the combined action of ordinary gauge rotations plus \eqref{defOmega} as a {\emph{spurious gauge transformation}}. Similarly, if $\Omega$ is also assumed to transform appropriately under global chiral rotations we are guaranteed that a spurious version of the global chiral transformations preserved by the Yukawa interactions remain conserved as well. (A generalization with spurious $d$-dimensional Lorentz and gauge invariance is also presented in Appendix \ref{sec:Lorentz}.)

From the defining property \eqref{defOmega} and by analogy with the theory of pions in QCD, we know that we can take $\Omega$ to be a special unitary matrix, a condition obviously consistent with the background value reported in \eqref{physback}. We will make heavy use of the convention $\Omega^\dagger\Omega=\Omega\Omega^\dagger=1$ in Section \ref{sec:radcorr}.

Besides being invariant under spurious chiral transformations, \eqref{Action} is also endowed with the same global symmetries of the standard regularized theory: it is invariant under $d$-dimensional translations, four-dimensional Lorentz transformations as well as rotations in the extra dimensions; moreover, it is symmetric under spurious versions of P and C, under which
\begin{equation}\label{spuriousCPP}
    \begin{split}
    A^A_\mu(x)&\xrightarrow[]{\text{P}}{\cal P}_\mu^\nu A^A_\nu({\cal P}x)\\
    \Psi_i(x)&\xrightarrow[]{\text{P}}\gamma^0\Psi_i({\cal P}x)\\
    \phi_a(x)&\xrightarrow[]{\text{P}}\phi_a({\cal P}x)\\
    \Omega_{ij}(x)&\xrightarrow[]{\text{P}}\Omega^*_{ji}({\cal P}x)\\
    T^A_{ij}&\xrightarrow[]{\text{P}}\overline{T}^A_{ij}\\
    \overline T^A_{ij}&\xrightarrow[]{{P}}{T}^A_{ij}\\
    [T_\phi^A]_{ab}&\xrightarrow[]{\text{P}}[T_\phi^A]_{ab}\\
    Y_{ij}^a&\xrightarrow[]{\text{P}}{Y_{ji}^a}^*
    \end{split}
       ~~~~~~~~~~~~~~~~~~
    \begin{split}
    A^A_\mu(x)&\xrightarrow[]{\text{C}}-A^A_\mu(x)\\
        \Psi_i(x)&\xrightarrow[]{\text{C}}C\overline\Psi_i^t(x)\\
    \phi_a(x)&\xrightarrow[]{\text{C}}\phi_a(x)\\
    \Omega_{ij}(x)&\xrightarrow[]{\text{C}}\Omega_{ij}^*(x)\\
    T^A_{ij}&\xrightarrow[]{\text{C}}{T_{ij}^A}^*\\
    \overline T_{ij}^A&\xrightarrow[]{\text{C}}{\overline{T}_{ij}^A}^*\\
    [T_\phi^A]_{ab}&\xrightarrow[]{\text{C}}[T_\phi^A]_{ab}^*\\
    Y^a_{ij}&\xrightarrow[]{\text{C}}{Y^a_{ij}}^*
\end{split}
\end{equation}
where ${\cal P}_\mu^0 x^\mu=x^0$ and ${\cal P}_\mu^{\nu\neq0} x^\mu=-x^{\nu\neq0}$. These relations implicitly involve a unitary rotation of the fields, which for brevity is not displayed. Note that a definition of the operator $C$ with the same properties of its four-dimensional version can be generalized to $d\neq4$ dimensions \cite{Belusca-Maito:2020ala}.

\section{Structure of the radiative corrections}
\label{sec:radcorr}

In this section we will see why working with \eqref{Action} is more convenient than with the standard theory in \eqref{LagTot}. The basic reason is that calculations carried out with \eqref{Action} are symmetric under spurious chiral rotations. This significantly constrains the structure of the radiative corrections and provides a precious guide throughout any computation.

In this paper we will adopt the background gauge formalism and a background-gauge invariant gauge-fixing \cite{Kluberg-Stern:1974nmx,Kluberg-Stern:1975ebk,Abbott:1981ke} 
\be\label{gf}
S_{\text{gf}}[a,A_c]=-\frac{1}{2}(D_{c}^\mu a_\mu)^A(D_{c}^\nu a_\nu)^A
\ee
where $D_{c}^\mu a_\mu=\partial^\mu a_\mu-i[A_{c}^\mu,a_\mu]$ denotes the covariant derivative of the quantum fluctuation $a_\mu=A_\mu-A_{c\,\mu}=a_\mu^AT^A$ with respect to the background field $A_{c\,\mu}=A_{c\,\mu}^AT^A$. Note that \eqref{gf} corresponds to a Feynman-’t
Hooft gauge for all (abelian and simple) group factors. 

The central observation of our paper is the following. The generating functional of 1-particle irreducible diagrams constructed with \eqref{Action} and \eqref{gf} is invariant under the spurious chiral symmetry. It can hence be split into two terms: the first involves amplitudes that are truly invariant and the second contains amplitudes that are invariant {\emph{only}} because of $\Omega$. The latter class is associated to the ``symmetry-violating" contributions characterizing the standard theory of Eq. \eqref{LagTot}. The explicit dependence on $\Omega$ thus allows us to straightforwardly isolate the offending terms, both the finite and singular parts. The spurious chiral-invariance is not only a very useful bookkeeping device, though, but in fact it also leads to practical advantages. One important advantage is in the extraction of the symmetry-restoring counterterms, which will be discussed in detail below. Another is that the ``symmetry-violating" terms may be determined by inspecting diagrams with external $\Omega$ lines, a procedure which may simplify their evaluation in some cases, as one for example is dispensed from having to deal with the different polarizations of the vector fields.~\footnote{We thank Sergey Sibiryakov for emphasizing this aspect.}

\subsection{Symmetry-restoring counterterms: general considerations}
\label{sec:grcgen}

The standard procedure used to identify the symmetry-restoring counterterms in a theory satisfying \eqref{D=0} consists in computing a naive one-particle irreducible (1PI) effective action with the standard bare action \eqref{LagTot}, and taking its symmetry variation; this gives the ``anomaly", which order by order in parturbation theory is a local functional. Alternatively one can directly compute the anomaly order by order via the insertion of appropriate local operators. Subsequently one needs to formally integrate the anomaly in order to find an appropriate set of symmetry-restoring counterterms whose variation precisely cancels the anomaly order by order in the $\hbar$ expansion.

In this subsection we show how the determination of the symmetry-restoring counterterms is considerably simplified by the introduction of $\Omega$. The basic object we are interested in is the 1PI effective action for the regularized theory $S^{\text{Reg}}$ given in \eqref{Action}, defined with the (background) gauge-invariant gauge-fixing \eqref{gf} and the associated ghost term $S_{\text{ghost}}$. Treating $\Omega$ as a non-dynamical external source, that quantity is defined as:
\be\label{preGamma}
e^{i\Gamma_{\text{inv}}[\xi_c,\Omega]}=\substack{{\text{LIM}}\\{d\to4}}\int_{\rm 1PI}{\cal D}\xi~e^{iS^{\text{Reg}}_{\text{tot}}[\xi,\xi_c,\Omega]+S_{\text{ct}}^{\text{Div}}[\xi+\xi_c,\Omega]+S_{\text{ct}}^{\text{Fin}}[\xi+\xi_c,\Omega]},
\ee
where $\xi$ collectively denotes all quantum degrees of freedom (including ghosts), $\xi_c$ their backgrounds, and
\be\label{Stot}
S^{\text{Reg}}_{\text{tot}}[\xi,\xi_c,\Omega]=S^{\text{Reg}}[\xi+\xi_c,\Omega]+S_{\text{gf}}[a,A_c]+S_{\text{ghost}}[\xi,\xi_c].
\ee
By $\substack{{\text{LIM}}\\{d\to4}}$ we indicate the limit in which all evanescent terms are dropped and $d\to4$, whereas the counterterms are defined as follows. The ``divergent" counterterms $S_{\text{ct}}^{\text{Div}}$ are introduced to absorb all (evanescent and non-evanescent) $1/(d-4)$ poles according to the $\overline{\text{MS}}$ renormalization scheme. Their determination is standard. The only novelty added by our formalism is that $S_{\text{ct}}^{\text{Div}}$ can be taken to be invariant under the spurious chiral symmetry. Indeed, $S^{\text{Reg}}_{\text{tot}}$ in \eqref{Stot} is symmetric under a backgound gauge transformation (combined with a linear field redefinition of the quantum field); as a result all divergences induced by it are chiral-invariant order by order, and this in turn guarantees that $S_{\text{ct}}^{\text{Div}}$ can be taken to be invariant as well. Analogously, finite counterterms $S_{\text{ct}}^{\text{Fin}}$ invariant under the spurious chiral symmetry can be added at will. The resulting expression $\Gamma_{\text{inv}}[\xi_c,\Omega]$ is therefore invariant by construction: a chiral transformation of the background fields $\xi_c$ gets exactly compensated by a transformation of the spurion $\Omega$. 

We define $S_{\text{ct}}^{\text{Fin}}$ such that chiral-invariance of \eqref{preGamma} is also ensured in the limit $\Omega={\mathbb 1}$, where \eqref{preGamma} reduces to the invariant 1PI effective action of the standard theory, i.e. $\Gamma_{\text{inv}}[\xi_c,{\mathbb 1}]$. Hence $S_{\text{ct}}^{\text{Fin}}[\xi_c,{\mathbb 1}]$ are the symmetry-restoring counterterms we alluded to below \eqref{D=0}. To identify them we follow an iterative procedure. Assume we have removed all divergences up to the $n$-th order in perturbation theory, but that chiral-invariance of the standard theory is only satisfied up to the order $n-1$. In other words, assume we have computed
\be\label{preGammaNOT}
e^{i\Gamma[\xi_c,\Omega]|_{(n)}}=\substack{{\text{LIM}}\\{d\to4}}\int_{\rm 1PI}{\cal D}\xi~e^{iS^{\text{Reg}}_{\text{tot}}[\xi,\xi_c,\Omega]+iS_{\text{ct}}^{\text{Div}}[\xi+\xi_c,\Omega]|_{(n)}+iS_{\text{ct}}^{\text{Fin}}[\xi+\xi_c,\Omega]|_{(n-1)}},
\ee
where $|_{(n)}$ indicates that the corresponding counterterm contains all loop effects up to the $n$-th order. Under our assumptions $\Gamma[\xi_c,\Omega]|_{(n)}$ is finite. Furthermore, $S_{\text{ct}}^{\text{Fin}}[\xi,\Omega]|_{(n-1)}$ ensures that $\Gamma[\xi_c,{\mathbb 1}]|_{(n-1)}$ be chiral-invariant, though in general $\Gamma[\xi_c,{\mathbb 1}]|_{(n)}$ is not. Our goal is to see how chiral-invariance of the standard theory is restored at order $n$, namely we want to find $S_{\text{ct}}^{\text{Fin}}[\xi,\Omega]|_{(n)}$.

In general we can split our functional \eqref{preGammaNOT} in a $\Omega$-dependent part 
\be
\Gamma_{\Omega}[\xi_c,\Omega]|_{(n)}
\ee
and a $\Omega$-independent part 
\be
\Gamma_{\slashed{\Omega}}[\xi_c]|_{(n)}\equiv\Gamma[\xi_c,\Omega]|_{(n)}-\Gamma_{\Omega}[\xi_c,\Omega]|_{(n)}. 
\ee
Because a chiral transformation cannot mix $\Omega$-dependent and $\Omega$-independent terms, both $\Gamma_{\Omega}|_{(n)}$ and $\Gamma_{\slashed{\Omega}}|_{(n)}$ must be separately invariant under the spurious chiral symmetry. An infinitesimal symmetry variation of the background fields leaves the latter completely invariant, i.e. 
\be
\delta_{\xi_c}(\Gamma_{\slashed{\Omega}}[\xi_c]|_{(n)})=0,
\ee
whereas generically acts non-trivially on $\Gamma_{\Omega}|_{(n)}$. In other words, $\Gamma_{\slashed{\Omega}}|_{(n)}$ is genuinely chiral-invariant while $\Gamma_{\Omega}|_{(n)}$ is only invariant because of $\Omega$. The quantity
\be\label{anomaly}
\delta_{\xi_c}(\Gamma_{\Omega}[\xi_c,{\mathbb 1}]|_{(n)})\equiv{\cal A}[\xi_c]|_{(n)}
\ee
corresponds to the chiral anomaly of the $n$-th order 1PI action of the standard theory. From this immediately follows that $\Gamma_{\Omega}[\xi_c,{\mathbb 1}]|_{(n)}$ must be local, since the anomaly is local and the variation $\delta_{\xi_c}$ cannot turn a non-local functional into a local one.

Now, the part of the symmetry-restoring counterterm of order $n$
\be
\Delta S_{\text{ct}}^{\text{Fin}}[\xi_c,{\mathbb 1}]|_{(n)}=S_{\text{ct}}^{\text{Fin}}[\xi_c,{\mathbb 1}]|_{(n)}- S_{\text{ct}}^{\text{Fin}}[\xi_c,{\mathbb 1}]|_{(n-1)}
\ee
is implicitly defined by
\be\label{defSct}
\delta_{\xi_c}(\Delta S_{\text{ct}}^{\text{Fin}}[\xi_c,{\mathbb 1}]|_{(n)})=-{\cal A}[\xi_c]|_{(n)}.
\ee
Comparing \eqref{anomaly} and \eqref{defSct} we see that the $n$-th order contribution to the symmetry variation of $S_{\text{ct}}^{\text{Fin}}[\xi_c,{\mathbb 1}]|_{(n)}$ must be the opposite of the variation of $\Gamma_{\Omega}[\xi_c,{\mathbb 1}]|_{(n)}$, see \eqref{anomaly}. This implies that the two functionals coincide up to a term that is invariant and $\Omega$-independent --- and hence by construction residing in $\Gamma_{\slashed{\Omega}}|_{(n)}$. Conventionally taking the $\Omega$-independent chiral-invariant term equal to zero, without loss of generality we can thus identify
\be\label{identification}
\Delta S_{\text{ct}}^{\text{Fin}}[\xi_c,{\mathbb 1}]|_{(n)}=-\Gamma_{\Omega}[\xi_c,{\mathbb 1}]|_{(n)}.
\ee
Our conclusion is therefore that the $n$-th order correction $\Delta S_{\text{ct}}^{\text{Fin}}[\xi_c,{\mathbb 1}]|_{(n)}$ necessary to restore chiral invariance in the standard theory can be obtained iteratively by computing (the opposite of) the finite, $\Omega$-dependent part of the $n$-th loop order 1PI functional \eqref{preGammaNOT} and sending $\Omega\to{\mathbb 1}$ at the very end. There is no need of taking an infinitesimal symmetry variation and subsequently integrating it.

Armed with the finite counterterm one can construct the chiral-invariant 1PI effective action of the standard theory at $n$-th order: $\Gamma_{\text{inv}}[\xi_c,{\mathbb 1}]|_{(n)}=\Gamma[\xi_c,{\mathbb 1}]|_{(n)}+\Delta S_{\text{ct}}^{\text{Fin}}[\xi_c,{\mathbb 1}]|_{(n)}$. Iterating the above procedure one obtains \eqref{preGamma} at the desired order in the perturbative expansion.

At 1-loop the analysis is particularly simple. In fact, because the classical anomaly is evanescent, its effect can appear only as a $1/(d-4)$ singular evanescent amplitude or as a finite amplitude resulting from the compensation of a $1/(d-4)$ pole and evanescent terms. As a result $\substack{{\text{LIM}}\\{d\to4}}\; S_{\text{ct}}^{\text{Div}}$ is genuinely symmetric (even without $\Omega$) and the 1-loop anomaly is manifestly local, as expected on general grounds.

\subsection{Symmetry-restoring counterterms: 1-loop calculation}
\label{sec:grc1}

As an application of our spurion trick we now derive $S_{\text{ct}}^{\text{Fin}}$ at the first non-trivial order using the identification \eqref{identification}.

Strictly speaking, $\Gamma_\Omega|_{(1)}$ should be evaluated with a space-time dependent $\Omega$. Yet, in our analysis we found it convenient to assume $\Omega=\Omega_0$ is an arbitrary constant unitary matrix. This way the fermionic propagator can be easily found for arbitrary $\Omega_0$,
\be
\frac{i}{p^2}\left[\gamma^{\bar\mu}p_{\bar\mu}+\left(\Omega_0 P_L+\Omega_0^\dagger P_R\right)\gamma^{\hat\mu}p_{\hat\mu}\right],
\ee
and $\Gamma[\xi_c,\Omega_0]|_{(1)}$ straightforwardly computed. However, with such a choice the covariant derivatives of $\Omega$ that appear in $\Gamma_\Omega$ do not display the $\partial_\mu\Omega$ term that would allow us to immediately identify the corresponding gauge-invariant operators. In practice, only a subset of the terms in $\Gamma_\Omega|_{(1)}$ depend explicitly on $\Omega_0$: the operators that are not invariant under the {\emph{chiral}} global symmetry when $\Omega_0\to{\mathbb 1}$. Nevertheless, identifying them turns out to be sufficient to fully reconstruct all symmetry-restoring counterterms because of the underlying spurious gauge-invariance enjoyed by $\Gamma|_{(1)}$. In fact, we can extract the symmetry-restoring counterterms as follows:
\begin{itemize}[leftmargin=1.53cm]
    \item[{\bf{Step 1:}}] First we compute the $\Omega_0$-dependent terms in $\Gamma_\Omega[\xi_c,\Omega_0]|_{(1)}$. For this, it is enough to compute the hard region of $\Gamma[\xi_c,\Omega_0]|_{(1)}$, because all the interactions of $\Omega$ are evanescent and therefore can only contribute if multiplied by the UV poles of the divergent integrals. In other words, $\Gamma_\Omega[\xi_c,\Omega_0]|_{(1)}\equiv\Gamma^{\mathrm{(hard)}}_\Omega[\xi_c,\Omega_0]|_{(1)}$.

    \item[{\bf{Step 2:}}] We then write a basis of counterterms $S_{\text{ct}}^{\text{Fin}}[\xi_c,\Omega]|_{(1)}$ invariant under the spurious chiral symmetry and the spurious P and C of \eqref{spuriousCPP}, with arbitrary coefficients, and evaluate the basis for $\Omega=\Omega_0$. 
    
    \item[{\bf{Step 3:}}] According to \eqref{identification}, the $\Omega_0$-dependent terms in  $\Gamma_\Omega[\xi_c,\Omega_0]|_{(1)}$ must exactly match (minus) those in $S_{\text{ct}}^{\text{Fin}}[\xi_c,\Omega_0]|_{(1)}$. Consistently, this is indeed the case: all the coefficients of the counterterm basis are uniquely determined. 
\end{itemize}

The 1-loop expression of $\Gamma_\Omega[\xi_c,\Omega_0]|_{(1)}$ is derived for definiteness assuming $c=0$ in \eqref{Action} and the Feynman-'t Hooft gauge of \eqref{gf}. The calculation was performed with {\tt Matchmakereft} \cite{Carmona:2021xtq}, in which the BMHV prescription has been implemented in a version that will be made public in the future.
$\Gamma_\Omega^{({\text{hard}})}[\xi_c,\Omega_0]|_{(1)}$ is a local object, and thus can be expressed in a basis of local operators, via a procedure completely parallel to the matching of UV poles during the renormalization of an effective field theory. The only difference is that  in the present case the coefficients are finite. Since {\tt Matchmakereft} is prepared to compute the hard region of a 1PI action to calculate the renormalization group equations, it is by construction well suited also to compute $\Gamma_\Omega^{({\text{hard}})}[\xi_c,\Omega_0]|_{(1)}$ automatically.

The list of relevant counterterms $S_{\text{ct}}^{\text{Fin}}[\xi_c,\Omega]|_{(1)}$ is discussed below and summarized in Tables \ref{Tab:1}, \ref{Tab:2}, \ref{Tab:3}, and Eq. \eqref{WZW}. Matching the two sets of $\Omega_0$-dependent operators, as explained in Step 3 above, we obtain the 1-loop coefficients reported in the right hand side of the tables. As a consistency check of our calculation we verified that exactly the same counterterms and the same coefficients are obtained via the standard method reviewed in the first paragraph of Section \ref{sec:grcgen}. Our results also reduce to those of \cite{Cornella:2022hkc} in the limit in which the scalar fields are decoupled. The explicit case of the Standard Model can be derived as a particular limit using the expressions for the generators and the Yukawa coupling collected in Appendix \ref{app:SM}.

In conclusion, we see two main concrete advantages of our approach. First, the symmetry-restoring counterterms are automatically found by computing a subset of 1-particle irreducible diagrams using \eqref{preGammaNOT}, there is no need to first evaluate the quantum anomaly and subsequently integrate it. 
Second, since we compute the effective action instead of its gauge variation, the number of diagrams is smaller. This is because in the calculation of the anomaly using the standard approach one would need to compute an insertion of the gauge variation of the classical action for each fermionic propagator, so that the number of diagrams to compute scales as the number of fermionic propagators in the amplitude.

%.... since the counterterms are directly extracted from the full $\Omega_0$-dependence, as opposed to looking at the first variation of the effective action, the number of diagrams to be computed can be much smaller than in the standard procedure. 

\subsubsection{Basis of counterterms}

The counterterms $S_{\text{ct}}^{\text{Fin}}[\xi_c,\Omega]|_{(1)}$ are invariants built out of the basic (four-dimensional) building blocks\footnote{To avoid cluttering we do not write the subscript $_c$ in the fields. It is understood from the general analysis of Section \ref{sec:grcgen} that these counterterms are expressed in terms of the backgound fields.}
\be\label{buildingblocks}
\Omega,A^A_{\bar\mu},\Psi_i,\overline{\Psi_i},\phi_a,D_{\bar\mu}, \gamma^{\bar\mu}
\ee
as well as the vertex spurions
\be\label{buildingblocks1}
T^A, \overline{T}^A, T_\phi^A, Y^a
\ee
appropriately contracted among each other with $\delta_{ij}$, $\delta_{AB}$, $\delta_{ab}$ from the fermion, gauge, scalar propagators. Note that in our convention the gauge couplings are included in the generators, so the number of \eqref{buildingblocks1} in a given operator effectively counts the number of coupling insertions.

The full list of the four-dimensional $\Omega$-dependent structures can be divided in four classes: ``Operators with only vectors and no Levi-Civita" (Table \ref{Tab:1}), ``Operators with scalars and no fermions" (Table \ref{Tab:2}), ``Operators with fermions" (Table \ref{Tab:3}), and finally ``Operators with the Levi-Civita tensor" (see Eq. \eqref{WZW}). All of them are invariant under the spurious chiral symmetry and also symmetric under the spurious P and C of \eqref{spuriousCPP}. It is important to emphasize that not all counterterms must be included in a given theory. To better appreciate this we distinguish among three classes of theories:
\begin{itemize}

\item {\it{Theories with chiral gauge symmetries.}} In this case all the counterterms in the tables are in general necessary (though some might identically vanish in a specific theory). This is the case of the Standard Model, which can be recovered with the help of Appendix \ref{app:SM}.
\item {\it{Theories with vector-like gauge symmetries and global chiral symmetries.}} Now all counterterms containing a covariant derivative of $\Omega$ vanish and only the two on the left of Table \ref{Tab:2} and the one on the right of Table \ref{Tab:3} can survive. The presence of a $\psi^2\phi$ term in particular implies that in the Standard Model we have a Yukawa-like counterterm proportional to the QCD and Yukawa couplings.
\item {\it{Theories with no chiral symmetries.}} Obviously, when no chiral symmetry (global nor local) is present one does not need to add any symmetry-restoring counterterm. $\Omega$ is not necessary in the first place. 
\end{itemize}

In identifying the structure of the finite counterterms one essentially follows the same rules used to construct the chiral Lagrangian for the pion matrix, whose role in the present context is played by $\Omega$. There are only a couple of important differences. First, the would-be kinetic term $\langle\partial_\mu\Omega^\dagger\partial^\mu\Omega\rangle$, where $\langle\,\dots\rangle$ indicates the matrix trace, does not appear at any order because of dimensional considerations. Hence, as opposed to what is done in the chiral Lagrangian, one cannot remove some of the higher-derivative interactions making field redefinitions of $\Omega$.

Second, here the underlying theory is assumed to be perturbative, and this fact constrains the type of operators that can be generated at each order in the perturbative expansion. The set of counterterms that are necessary is therefore a very small subgroup of all operators compatible with the spurious symmetries. To start, at 1-loop order and with $c=0$ the counterterms contain at most a trace over the flavor indices; double-trace operators will be generated starting at two loops. Furthermore, an inspection of the topology of the 1PI diagrams immediately suggests which structures can be relevant in each class. Let us discuss this more explicitly.

``Operators with only vectors and no Levi-Civita" are generated by a single fermionic loop with no virtual bosons. Hence, no explicit dependence on the Yukawa coupling nor the gauge generators can appear. Taking this into account the complete set of invariant $\Omega$-dependent single-trace dimension-4 operators without \eqref{buildingblocks1} is shown in Table \ref{Tab:1}. The 1-loop coefficients of the counterterms, in units of $1/(16\pi^2)$, is shown for each of the operators in the second column of Table \ref{Tab:1}. We note that a linear combination of those operators is also gauge-invariant (up to a total derivative of some vector $T^\mu$) in the limit $\Omega\to{\mathbb 1}$, since
\begin{align}\label{gaugeinvcomb}
    &\langle L_{\bar\mu\bar\nu}L^{\bar\mu\bar\nu}+R_{\bar\mu\bar\nu}R^{\bar\mu\bar\nu}\rangle+\partial_\mu T^\mu=\\\no
    &\langle
    R_{\mu\nu}\Omega^\dagger L^{\mu\nu}\Omega-{i}\left(R_{\mu\nu}D^\mu\Omega^\dagger D^\nu\Omega+L_{\mu\nu}D^\mu\Omega D^\nu\Omega^\dagger\right)+D_\mu D_\nu\Omega D^\mu D^\nu\Omega^\dagger-D_\mu D^\mu\Omega D_\nu D^\nu\Omega^\dagger\rangle_{\Omega={\mathbb 1}},
\end{align}
where we defined $L_{\bar\mu}=A^A_{\bar\mu} T^A_L$ and $L_{\bar\mu\bar\nu}=i[D_{\bar\mu},D_{\bar\nu}]$, and analogously for $R_{\bar\mu}$. The coefficient of the combination \eqref{gaugeinvcomb} therefore depends on the renormalization scheme.

Similarly, ``Operators with only scalars" are generated by a fermionic loop with external $\phi$ or $\Omega$. Hence they can only depend on the external $\phi$ via the gauge-covariant combination
\be
\Phi_{ij}=Y^a_{ij}\phi_a.
\ee
All hermitian, single-trace and dimension-4 operators involving $\Phi$ and compatible with our symmetries have non-vanishing coefficients at 1-loop, see Table \ref{Tab:2}. 

Analogous considerations demonstrate that only a very limited number of ``Operators with fermions" is generated. The topology of the 1-loop diagrams evidently selects what string of couplings and spurions is chained within the external $\overline{\Psi}$ and $\Psi$ lines. The counterterms induced in our 1-loop analysis are shown in Table \ref{Tab:3} along with their coefficients.

%%%%%%%%%%%%%%%%%%%%%% Pure gauge (no Levi-Civita) %%%%%%%%%%%%%%%%%%%%%%
\begin{table}[h!]
\centering
\begin{tabular}{|l|l|}
\hline
\multicolumn{2}{|>{\columncolor[gray]{0.90}}c|}{$\boldsymbol{D^4}$}
\\ 
\hline
$\langle L_{\bar\mu\bar\nu}\Omega R^{\bar\mu\bar\nu}\Omega^\dagger\rangle$ & $0$
\tabularnewline
$i \langle L_{\bar\mu\bar\nu}D^{\bar\mu}\Omega D^{\bar\nu}\Omega^\dagger+R_{\bar\mu\bar\nu}D^{\bar\mu}\Omega^\dagger D^{\bar\nu}\Omega\rangle$ & $-\frac{1}{2}$
\tabularnewline
$\langle D_{\bar\mu}\Omega D^{\bar\mu}\Omega^\dagger D_{\bar\nu}\Omega D^{\bar\nu}\Omega^\dagger+D_{\bar\mu}\Omega^\dagger D^{\bar\mu}\Omega D_{\bar\nu}\Omega^\dagger D^{\bar\nu}\Omega \rangle$ & $-\frac{1}{6}$
\tabularnewline
$\langle D_{\bar\mu}\Omega D_{\bar\nu}\Omega^\dagger D^{\bar\mu}\Omega D^{\bar\nu}\Omega^\dagger\rangle$ & $+\frac{1}{12}$
\tabularnewline
$\langle D_{\bar\mu} D^{\bar\mu}\Omega D_{\bar\nu} D^{\bar\nu}\Omega^\dagger\rangle$ & $0$
\tabularnewline
$\langle D_{\bar\mu} D_{\bar\nu}\Omega D^{\bar\mu} D^{\bar\nu}\Omega^\dagger\rangle$ & $+\frac{1}{6}$
\\
[5pt]
\hline
\end{tabular}
\caption{Counterterms and 1-loop coefficients (in units of ${1}/({16\pi^2})$) in the class ``Operators with only vectors and no Levi-Civita". The symbol $\langle\,\dots\rangle$ stands for the trace in the fermionic indices. We take $c=0$ in \eqref{Action} and work in the Feynman-'t Hooft gauge \eqref{gf}.\label{Tab:1}}
\end{table}
%%%%%%%%%%%%%%%%%%%%%%%%%%%%%%%%%%%%%%%%%%%%%%%%%%%%%%%%%%%

%%%%%%%%%%%%%%%%%%%%%% Pure scalars %%%%%%%%%%%%%%%%%%%%%%
\begin{table}[h!]
\begin{centering}
\begin{tabular}{|l|l|l|l|}
\hline
\multicolumn{2}{|c|}{\cellcolor[gray]{0.90}$\boldsymbol{\phi^4}$}
&
\multicolumn{2}{c|}{\cellcolor[gray]{0.90}$\boldsymbol{\phi^2 D}$}
\\ 
\hline 
$\langle(\Phi\Omega^\dagger)^4\rangle+{\text{h.c.}}$ & $-\frac{1}{12}$
&
$\langle\Phi D_{\bar\mu}\Omega^\dagger\Phi D^{\bar\mu}\Omega^\dagger\rangle+{\text{h.c.}}$ & $+\frac{1}{3}$
\tabularnewline
$\langle(\Phi\Omega^\dagger)^2\Phi\Phi^\dagger\rangle+{\text{h.c.}}$ & $-\frac{2}{3}$ 
&
$\langle(\Phi\Omega^\dagger)^2D_{\bar\mu}\Omega D^{\bar\mu}\Omega^\dagger\rangle+{\text{h.c.}}$ & $-\frac{1}{3}$
\tabularnewline
$$ & $$ 
& $\langle\Phi\Phi^\dagger D_{\bar\mu}\Omega D^{\bar\mu}\Omega^\dagger+\Phi^\dagger\Phi D_{\bar\mu}\Omega^\dagger D^{\bar\mu}\Omega\rangle$ & $-\frac{1}{3}$
\tabularnewline
$$ & $$ 
& $\langle\Phi \Omega^\dagger D_{\bar\mu}\Omega\Phi^\dagger \Omega D^{\bar\mu}\Omega^\dagger \rangle$ & $+\frac{1}{3}$
\tabularnewline
$$ & $$ 
& $\langle D_{\bar\mu}\Phi D^{\bar\mu}\Omega^\dagger \Phi\Omega^\dagger+D_{\bar\mu}\Phi \Omega^\dagger\Phi D^{\bar\mu}\Omega^\dagger\rangle+{\text{h.c.}}$ & $+\frac{2}{3}$
\tabularnewline
$$ & $$ 
& $\langle D_{\bar\mu}\Phi \Omega^\dagger D^{\bar\mu}\Phi\Omega^\dagger\rangle+{\text{h.c.}}$ & $+\frac{1}{6}$
\tabularnewline
$$ & $$ 
& $\langle \Phi \overleftrightarrow{D}_{\bar\mu}\Phi^\dagger\Omega D^{\bar\mu}\Omega^\dagger+\Phi^\dagger\overleftrightarrow{D}_{\bar\mu}\Phi\Omega^\dagger D^{\bar\mu}\Omega\rangle$ & $+\frac{1}{3}$
\tabularnewline
[8pt]
\hline
\end{tabular}
\caption{Same as in Table \ref{Tab:1} for the class ``Operators with scalars and no fermions".\label{Tab:2}}
\end{centering}
\end{table}
%%%%%%%%%%%%%%%%%%%%%%%%%%%%%%%%%%%%%%%%%%%%%%%%%%%%%%%%%%%

%%%%%%%%%%%%%%%%%%%%%% Fermions %%%%%%%%%%%%%%%%%%%%%%
\begin{table}[h!]
\begin{centering}
\begin{tabular}{|l|l|l|l|}
\hline
\multicolumn{2}{|c|}{\cellcolor[gray]{0.90}$\boldsymbol{\psi^2D}$}
&
\multicolumn{2}{c|}{\cellcolor[gray]{0.90}$\boldsymbol{\psi^2 \phi}$}
\\ 
\hline 
$\overline\Psi\gamma^{\bar\mu}T_L^A\Omega iD_{\bar\mu}\Omega^\dagger T_L^AP_L\Psi+{\text{P.c.}}$ & $1$
&
$[\overline\Psi T_R^A\Omega^\dagger \Phi \Omega^\dagger T_L^AP_L\Psi+{\text{P.c.}}]+{\text{h.c.}}$ & $-2$
\tabularnewline
$\overline\Psi \gamma^{\bar\mu}Y^a\Omega^\dagger iD_{\bar\mu}\Omega [Y^a]^\dagger P_L\Psi+{\text{P.c.}}$ & $\frac{1}{2}$ 
&
$$ & $$
\tabularnewline
[5pt]
\hline
\end{tabular}
\caption{Same as in Table \ref{Tab:1} for the class  ``Operators with fermions". By ${\text{P.c.}}$ we mean ``parity conjugate".\label{Tab:3}}
\end{centering}
\end{table}
%%%%%%%%%%%%%%%%%%%%%%%%%%%%%%%%%%%%%%%%%%%%%%%%%%%%%%%%%%%

The set ``Operators with the Levi-Civita tensor" is special. As well known, not all $\Omega$-dependent terms in the chiral Lagrangian can be written as invariant operators of a four-dimensional functional. There is only one exception: the Wess-Zumino-Witten term \cite{Wess:1971yu,Witten:1983tw}. In a theory satisfying \eqref{D=0}, the gauged version of this term can be written as\footnote{The form of the gauged Wess-Zumino-Witten term originally presented in \cite{Witten:1983tw} was not completely correct. We use the corrected version subsequently found in \cite{Chou:1983qy,Kawai:1984mx,Alvarez-Gaume:1984zlq} (see also \cite{Scherer:2002tk}).} 
\be\label{WZW}
\left.S_{\text{ct}}^{\rm Fin}[\xi,\Omega]\right|_{\text{WZW}}=\frac{n}{48\pi^2}\left\{\int \dd^4x~\epsilon^{\bar\mu\bar\nu\bar\alpha\bar\beta}\,Z_{\bar\mu\bar\nu\bar\alpha\bar\beta}+\dots\right\},
\ee
where
\begin{align}
Z_{\bar\mu\bar\nu\bar\alpha\bar\beta}
=&\,\langle-\Omega^\dagger\partial_{\bar\mu} L_{\bar\nu} L_{\bar\alpha}\Omega R_{\bar\beta}+\Omega\partial_{\bar\mu} R_{\bar\nu} R_{\bar\alpha}\Omega^\dagger L_{\bar\beta}\\\no
-&\partial_{\bar\mu} R_{\bar\nu}\Omega^\dagger L_{\bar\alpha}\Omega R_{\bar\beta}+\partial_{\bar\mu} L_{\bar\nu}\Omega R_{\bar\alpha}\Omega^\dagger L_{\bar\beta}\\\no
+&i\Omega^\dagger L_{\bar\mu} L_{\bar\nu} L_{\bar\alpha}\Omega R_{\bar\beta}-i\Omega R_{\bar\mu} R_{\bar\nu} R_{\bar\alpha}\Omega^\dagger L_{\bar\beta}\\\no
+&\frac{i}{2}\Omega^\dagger L_{\bar\mu}\Omega R_{\bar\nu} \Omega^\dagger L_{\bar\alpha}\Omega R_{\bar\beta}+{\cal O}(\partial\Omega)\rangle,
\end{align}
with $\epsilon^{0123}=-1$. The dots in \eqref{WZW} indicate terms that cannot be written as a four-dimensional functional, and that are proportional to the derivatives of the Goldstones. They vanish when setting $\Omega=1$ and are therefore not relevant to us.

Thanks to the spurious chiral-invariance we see that the $\epsilon^{\bar\mu\bar\nu\bar\alpha\bar\beta}$-dependent finite counterterm is fully determined by a unique coefficient $n$. Our 1-loop computation gives 
\be\label{n=1}
n=1,
\ee
a result that agrees with the earlier calculation of \cite{ Cornella:2022hkc}. Importantly, $n$ must be an integer in order for \eqref{WZW} to describe a consistent gauge-invariant four-dimensional theory \cite{Witten:1983tw}. This implies that \eqref{n=1} cannot receive perturbative corrections. The result $n=1$ is therefore exact at all orders in perturbation theory.

To conclude we stress that while the focus of this subsection is on $S_{\text{ct}}^{\text{Fin}}$, our spurious symmetries severely constrain the counterterms appearing in $S_{\text{ct}}^{\text{Div}}$ as well. Because the divergent counterterms in general include evanescent contributions, in identifying the list of operators in $S_{\text{ct}}^{\text{Div}}$ the set of building blocks \eqref{buildingblocks} should be extended to include also the evanescent counterparts.

\section{Conclusions}
\label{sec:conclusion}

In concrete quantum field theory calculations it is always preferable to adopt a regularization scheme that respects as many symmetries as possible. This way quantum corrections are severely constrained and our efforts can be guided by symmetry considerations and be subject to powerful crosschecks. Unfortunately, a consistent treatment of the $\gamma_5$ matrix in Dimensional Regularization results in an explicit breaking of all chiral symmetries, like those defining the Standard Model gauge symmetry. 

In this paper we emphasized that the introduction of appropriate auxiliary Goldstone fields allows to formally restore the chiral symmetry in dimensionally-regularized theories with arbitrary gauge group. If combined with the background gauge technique, chiral-invariance is therefore retained at all steps. This has the practical advantage that the structure of the radiative corrections becomes more transparent. The additional, seemingly unstructured counterterms appearing in Dim-Reg simply emerge because a chiral dimensionally regularized theory secretly has more fields than its four-dimensional cousin, i.e. the non-dynamical ``Goldstone spurions". Once one explicitates those fields, order is re-established.

A straightforward consequence of our formalism is that the symmetry-restoring counterterms are automatically delivered by the computation of 1PI diagrams, without the need of performing operator insertions and a subsequent ``integration" of the chiral anomaly, as instead done with the more standard methods. As a concrete application, we carried out a complete 1-loop calculation of the finite four-dimensional counterterms necessary to restore all (non-anomalous) chiral symmetries in renormalizable gauge theories with scalars and Dirac fermions, the Standard Model being a particular example. A sharp implication of our formalism is that the finite counterterms proportional to the Levi-Civita tensor are 1-loop exact.

It would be interesting to apply our formalism to more general theories, like for example the Standard Model effective field theory, as well as to go beyond the 1-loop approximation. In this latter case, restoring also full $d$-dimensional Lorentz invariance might turn out to be especially convenient.

\section*{Acknowledgments}

We would like to thank C. Cornella for discussions in the early stages of the project and S. Sibiryakov for renewing our interest on the topic. The work of LV was partly supported by the Italian MIUR under contract 202289JEW4 (Flavors: dark and intense), the Iniziativa Specifica ``Physics at the Energy, Intensity, and Astroparticle Frontiers" (APINE) of Istituto Nazionale di Fisica Nucleare (INFN), and the European Union’s Horizon 2020 research and innovation programme under the Marie Sklodowska-Curie grant agreement No 860881-HIDDeN. 
The work was also partially supported by grant FPU19/06813, an Arqus mobility grant, by MICIU/AEI/10.13039/501100011033 
and FEDER/UE
(grant PID2022-139466NB-C21), 
the European Union’s Horizon 2020 research and innovation programme under the Marie Sklodowska-Curie
grant agreement n. 101086085 – ASYMMETRY by the INFN Iniziativa Specifica APINE, and by the Italian MUR Departments of Excellence grant 2023-2027 ”Quantum Frontiers”.

\begin{appendices}

\section{The Standard Model case}
\label{app:SM}
%\addcontentsline{toc}{section}{\protect\numberline{}Appendix: the Standard Model case}

The Standard Model is a particular case of the theories considered in the present paper. In this appendix we present the explicit expressions for all the couplings and tensors necessary to compute our counterterms in such a theory.

The gauge group is $\mathrm{SU}(3)\times\mathrm{SU}(2)\times\mathrm{U}(1)$ and the vectors are embedded in a $12$-dimensional field. We can describe the particle content using an eight-dimensional fermionic multiplet $\Psi_i$ (with $i=1,\dots,8$) and a four-component scalar multiplet $\phi_a$ (with $a=1,\dots,4$) given by:
\begin{equation}
    \Psi=\begin{pmatrix}
       u\\d\\ \nu\\e\\
    \end{pmatrix},\qquad
    H=\frac{1}{\sqrt2}\begin{pmatrix}
        \phi_3+i\phi_4\\
        \phi_1+i\phi_2
    \end{pmatrix}.
\end{equation}
The left-handed components of the up-type ($u$) and the down-type ($d$) quarks form the familiar quark doublet $q_L=(u_L\ d_L)$, and similarly for the left-handed leptons $\ell_L=(\nu_L\ e_L)$. The right-handed neutrinos $\nu_R$ are included to keep the notation symmetric; they eventually decouple because neutral under $\mathrm{SU}(3)\times\mathrm{SU}(2)\times\mathrm{U}(1)$ and because their Yukawa interactions are assumed to be absent. The fermion family index is suppressed for simplicity throughout this appendix. Summation over the three Standard Model generations can be straightforwardly taken into account.

The Higgs doublet $H$ is taken to have hypercharge $+1/2$. The (opposite of the) Yukawa Lagrangian thus reads
\begin{equation}
-\mathcal{L}^{\mathrm{(SM)}}_{\mathrm{Yuk}}= 
\overline{q_L}y_u u_R \widetilde{H}
+\overline{q_L}y_d d_R H+\overline{\ell_L}y_e e_R H
+\mathrm{h.c.},
\end{equation}
with all gauge and flavor indices left as understood and $\widetilde{H}=i\sigma^2H^*$.

The ${\text{SU(3)}}$ gauge generators of the left(right)-handed fermion representations are
\begin{equation}\no
\left.T_{L}\right|_{\text{SU(3)}}=\left.T_{R}\right|_{\text{SU(3)}}=\begin{pmatrix}
        \frac12\lambda&&&\\
        &\frac12\lambda&&\\
        &&0&\\
        &&&0
    \end{pmatrix},
\end{equation}
with $\lambda$ denoting the Gell-Mann matrices (empty space indicates vanishing components). Similarly, the generators for the fermions of the electroweak ${\text{SU(2)}}\times{\text{U(1)}}$ gauge group are
\begin{align}\no
    &\left.T_{L}^{\left\{1,2,3\right\}}\right|_{\text{SU(2)}}=\left\{
    \frac12\begin{pmatrix}
        &{\mathbb 1}&&\\
        {\mathbb 1}&&&\\
        &&&1\\
        &&1&
    \end{pmatrix},
    \frac12\begin{pmatrix}
        &-i{\mathbb 1}&&\\
        +i{\mathbb 1}&&&\\
        &&&-i\\
        &&i&
    \end{pmatrix},
    \frac12\begin{pmatrix}
        {\mathbb 1}&&&\\
        &-{\mathbb 1}&&\\
        &&1&\\
        &&&-1
    \end{pmatrix}
    \right\},\\\no
    &\left.T_{R}^{\left\{1,2,3\right\}}\right|_{\text{SU(2)}}={\mathbb 0}
\end{align}
and
\begin{equation}
    \left.T_{L}\right|_{\text{U(1)}}=\begin{pmatrix}
        \frac16{\mathbb 1}&&&\\
        &\frac16{\mathbb 1}&&\\
        &&-\frac12&\\
        &&&-\frac12
    \end{pmatrix},\qquad
    \left.T_{R}\right|_{\text{U(1)}}=\begin{pmatrix}
        \frac23{\mathbb 1}&&&\\
        &-\frac13{\mathbb 1}&&\\
        &&0&\\
        &&&-1
    \end{pmatrix},
\end{equation}
with ${\mathbb 1}$ the identity matrix in three dimensions and ${\mathbb 0}$ the vanishing matrix in eight dimensions. The ${\text{SU(2)}}\times{\text{U(1)}}$ generators $T_\phi$ for the scalars $\phi_a$ can be written as:
\begin{align}\no
    &\left.T_{\phi}^{\left\{1,2,3\right\}}\right|_{\text{SU(2)}}=\left\{
    \frac12\begin{pmatrix}
        &&&i\\
        &&-i&\\
        &i&&\\
        -i&&&
    \end{pmatrix},
    \frac12\begin{pmatrix}
        &&i&\\
        &&&i\\
        -i&&&\\
        &-i&&
    \end{pmatrix},
    \frac12\begin{pmatrix}
        &-i&&\\
        i&&&\\
        &&&i\\
        &&-i&
    \end{pmatrix}
    \right\},\\
    &\left.T_{\phi}\right|_{\text{U(1)}}=\frac12\begin{pmatrix}
        &i&&\\
        -i&&&\\
        &&&i\\
        &&-i&
    \end{pmatrix}.
\end{align}

Finally, the Yukawa couplings $Y^a$ acquire the following form:
\begin{align}
Y^1=\begin{pmatrix}
+y_u{\mathbb 1}&&&\\
&+y_d{\mathbb 1}&&\\
&&0&\\
&&&+y_e
\end{pmatrix},~~~&~~~ Y^2=\begin{pmatrix}
-i y_u{\mathbb 1}&&&\\
&+i y_d{\mathbb 1}&&\\
&&0&\\
&&&+i y_e
\end{pmatrix},\nonumber\\
Y^3=\begin{pmatrix}
&+y_d{\mathbb 1}&&\\
-y_u{\mathbb 1}&&&\\
&&&+y_e\\
&&0&
\end{pmatrix},~~~&~~~
Y^4=
\begin{pmatrix}
&+iy_d{\mathbb 1}&&\\
+iy_u{\mathbb 1}&&&\\
&&&+i y_e\\
&&0&
\end{pmatrix}.
\end{align}

\section{d-dimensional Lorentz and gauge invariance}
\label{sec:Lorentz}

The formalism we have invoked in Section \ref{sec:spurion} is familiar from the study of non-linearly realized symmetries. The original global symmetry $G=SU(N)_L\times SU(N)_R$ of the fermionic kinetic term is broken by $\gamma^{\hat\mu}$ down to its diagonal subgroup of vector-like transformations $H=SU(N)_V$. Formally, we can view $\gamma^{\hat\mu}$ as a $G$-tensor invariant under $H$: $\gamma^{\hat\mu}\to h\gamma^{\hat\mu}h^\dagger=\gamma^{\hat\mu}$. To restore the complete $G$ we must introduce Nambu-Goldstone bosons $\Pi$ and a coset representative $\sqrt\Omega(\Pi)$ transforming as
$$
\sqrt\Omega\to L\sqrt\Omega h^\dagger(\Pi)=h(\Pi)\sqrt\Omega R^\dagger,
$$
with $L\in SU(N)_L$ and $R\in SU(N)_R$. Then we can define 
\be\label{sigma}
\sigma\equiv P_L\sqrt\Omega^\dagger+P_R\sqrt\Omega, 
\ee
such that $\sigma\to h(\Pi)\sigma U^\dagger= \overline{U}\sigma h^\dagger(\Pi)$, and use it to construct the $G$-covariant object 
\be
\sigma\gamma^{\hat\mu}\sigma=\gamma^{\hat\mu}(P_L\Omega^\dagger+P_R\Omega)\to \overline U\sigma\gamma^{\hat\mu}\sigma U^\dagger.
\ee
On the other hand $\sigma\gamma^{\bar\mu}\sigma=\gamma^{\bar\mu}$ remains unchanged. This argument explains the $\Omega$-dependence of \eqref{Action}.

We can introduce additional auxiliary fields to restore full $d$-dimensional Lorentz invariance. In such case the symmetry breaking pattern is $SO(1,d-1)\to SO(1,3)\times SO(d-4)$. What breaks the original $d$-dimensional Lorentz invariance is $\gamma_5$, which is a singlet only under the four-dimensional subgroup, namely $\gamma_5\to S(\Lambda_4)\gamma_5S^{-1}(\Lambda_4)=\gamma_5$, where $S(\Lambda_4)$ denotes the spinorial representation and $\Lambda_4\in SO(1,3)$. On the other hand $\gamma^{\mu}=\Lambda^\mu_{~\nu}S(\Lambda)\gamma^{\nu}S^{-1}(\Lambda)$, as usual, with $S(\Lambda)$ the spinorial representation associated to $\Lambda\in SO(1,d-1)$. Introducing a new set of Nambu-Goldstone bosons $\Pi'$ and a new coset representative 
\be
\Omega'\to \Lambda\Omega' \Lambda_4^{-1}(\Pi') 
\ee
we can build a $G$-covariant version of $\gamma_5$:
\be
\Gamma_5\equiv S(\Omega')\gamma_5S^{-1}(\Omega')\to S(\Lambda)\Gamma_5 S^{-1}(\Lambda).
\ee
 The Clifford algebra implies $S^\dagger\gamma^0=\gamma^0 S^{-1}$. Hence, by construction we have $
\gamma^0\Gamma_5^\dagger\gamma^0=-\Gamma_5$. Because $\Gamma_5^2=1$, we can define a Lorentz-covariant version of the chirality projectors 
\begin{align}
{\cal P}_L\equiv\frac12(1-\Gamma_5)~~~~~~~
{\cal P}_R\equiv\frac12(1+\Gamma_5)
\end{align}
and of \eqref{sigma}: 
\be
\Sigma\equiv{\cal P}_L\sqrt\Omega^\dagger+{\cal P}_R\sqrt\Omega. 
\ee
These satisfy $\gamma^0{\cal P}^\dagger_L\gamma^0={\cal P}_R$, and similarly $\gamma^0{\cal P}^\dagger_R\gamma^0={\cal P}_L$, as well as $\gamma^0\Sigma^\dagger\gamma^0=\Sigma$. Importantly, because the symmetry we are restoring is only global, the new Goldstones can be taken to be space-time independent. This allows us to treat ${\cal P}_{L,R}$ as a constant field.

With our new machinery ${\cal P}_L\Psi$ and ${\cal P}_R\Psi$ are $d$-dimensional Lorentz spinors. We postulate they transform under the chiral gauge symmetry as 
$$
{\cal P}_L\Psi\to e^{i\alpha^AT_L^A}{\cal P}_L\Psi,~~~~~~~~{\cal P}_R\Psi\to e^{i\alpha^AT_R^A}{\cal P}_R\Psi. 
$$
The covariant derivatives read\footnote{Here it is crucial that ${\cal P}_{L,R}$ are constant fields.} 
$$
{\cal D}_\mu\Psi\equiv\left(\partial_\mu-iA^A_\mu{\cal T}^A\right)\Psi,~~~~~
{\cal D}_\mu\overline\Psi\equiv\partial_\mu\overline\Psi+iA^A_\mu\overline\Psi\,\overline{\cal T}^A,
$$
with ${\cal T}^A={\cal P}_LT^A_L+{\cal P}_RT^A_R$  and $\overline{\cal T}^A={\cal P}_RT^A_L+{\cal P}_LT^A_R$ Lorentz-covariant space-time independent gauge generators. An incarnation of \eqref{Action} that satisfies (a spurious) global $d$-dimensional Lorentz invariance as well as gauge-invariance may finally be written as
\begin{align}\label{Action1}
%S^{\text{Reg}}[\xi,\Omega,\Omega']&=
\int \dd^dx\left\{{\cal L}_{\text{Bos}}^{\text{Reg}}+\left[\frac12\overline{\Psi}_j\Sigma\gamma^{\mu}\Sigma i{\cal D}_\mu\Psi_j+Y^a_{ij}\overline{\Psi_i}{\cal P}_R\Psi_j\phi_a+{\rm h.c.}\right]\right\}.
\end{align}
This expression reduces to \eqref{Action} when $\Omega'$ becomes the identity matrix, i.e. when ${\cal P}_{L,R}\to P_{L,R}$.

Intuitively, a regularization that is invariant under the full $d$-dimensional Lorentz group should be more constraining than one, like \eqref{Action}, that realizes explicitly only its four-dimensional subgroup. A manifestation of this tendency is that dimension-3 counterterms like $\overline\Psi\Omega P_R\Psi$, in principle allowed by the spurious symmetries of Section \ref{sec:spurion}, in the formulation \eqref{Action1} cannot occur at any order in perturbation theory. Indeed, such an hypothetical operator should be of the form $\overline\Psi O\Psi$, with $O$ an object constructed out of $\Sigma$, $\gamma^\mu$, $\Gamma_5$, and the couplings \eqref{buildingblocks1}, and no derivatives; but it is easy to see that there exists no structure of that form that is simultaneously gauge and Lorentz covariant.

The higher degree of constraining power is however partially compensated by the fact that \eqref{Action1} has more vertices involving non-dynamical Goldstones. In the calculation presented in Section \ref{sec:grc1} the formalism used in Sections \ref{sec:reg} and \ref{sec:radcorr} seemed to us more convenient. Yet, this may not be the case in more involved calculations.

\end{appendices}

\bibliographystyle{JHEP}
\bibliography{bibliography}

\end{document}